# ETP-Mine: An Efficient Method for Mining Transitional Patterns

B. Kiran Kumar[1] and A. Bhaskar[2]

[1]Department of M.C.A., Kakatiya Institute of Technology & Science, A.P., INDIA.
kirankumar.bejjanki@gmail.com
[2]Department of M.C.A., Kakatiya Institute of Technology & Science, A.P., INDIA.
bhaskar_adepu@yahoo.com

***ABSTRACT***

*A Transaction database contains a set of transactions along with items and their associated timestamps. Transitional patterns are the patterns which specify the dynamic behavior of frequent patterns in a transaction database. To discover transitional patterns and their significant milestones, first we have to extract all frequent patterns and their supports using any frequent pattern generation algorithm. These frequent patterns are used in the generation of transitional patterns. The existing algorithm (TP-Mine) generates frequent patterns, some of which cannot be used in generation of transitional patterns. In this paper, we propose a modification to the existing algorithm, which prunes the candidate items to be used in the generation of frequent patterns. This method drastically reduces the number of frequent patterns which are used in discovering transitional patterns. Extensive simulation test is done to evaluate the proposed method.*

## 1. INTRODUCTION

Frequent pattern mining was extensively studied in the area of data mining research. These frequent patterns are used in many different data mining tasks like Association rule mining [2] [3] [17], sequential pattern mining [4], structured pattern mining [10], correlation mining [6], and associative classification [12]. Many algorithms were proposed to find frequent patterns in transaction databases namely Apriori [3], FP-growth [11], and Eclat [23]. But these methods generate large number of patterns when the minimum support threshold is low. Out of these patterns most are not required in data mining tasks. To avoid these useless or redundant patterns, new types of patterns were introduced, namely maximal frequent itemsets [1] [5] [8], and closed frequent itemsets [15].

All the above methods do not consider time stamps associated with the transactions. So, the above methods do not reveal the dynamic behavior of the patterns. For example, in an electronics shop database, the sales for Air conditioners will be high during the summer season and the sales will be low in other seasons. If we consider total transactions, Air conditioners will become frequent, but these are frequent in only summer season. To capture the dynamic behavior of the patterns, new type patterns were introduced, called Transitional patterns [19] [20]. Transitional patterns can be both positive and negative. Positive transitional patterns increase their frequency at some time stamp. Negative transitional patterns decrease their frequency at some time stamp. Significant milestone of the transitional pattern is the time point at which the frequency of the transitional pattern changes most frequently. TP-Mine algorithm [20] was proposed to discover the transitional patterns and their significant milestones.

The TP-Mine algorithm contains two major steps. In the first step all the frequent patterns are generated from the transaction database. In the second step these frequent patterns are used in generating transitional patterns and their significant milestones within the given milestone range that the user is interested. But, some of the frequent patterns generated in the first step are not used in the second step.

In this paper we are proposing a modification to the TP-Mine algorithm to eliminate the useless frequent patterns generated in the first step. Thereby we can reduce the number of computations required in generating transitional patterns.

The rest of the paper is organized as follows. Section 2 describe relates work. Section 3 describes preliminaries and definitions used in TP-Mine algorithm. Section 4 describes TP-Mine Algorithm. In section 5, we describe the modified TP-Mine (ETP-Mine) algorithm. Section 6 presents experimental results of our work. Finally, in section 7 we conclude our work.

## 2. RELATED WORK

The Apriori [3] algorithm proposed by Agrawal and Srikanth observed an interesting downward closure property, called Apriori property i.e., "*Every subset of the frequent itemset is frequent*". This implies that frequent itemsets can be mined by first scanning the database to find frequent 1-itemsets, then using the frequent 1-itemsets to generate candidate 2-itemsets, and check against the database to obtain the frequent 2-itemsets. This process iterates until no more frequent k-itemsets can be generated for some k.

The problem with Apriori is that it generates too many 2-itemsets that are not frequent. Direct Hashing and Pruning (DHP) algorithm [13], reduces the size of candidate set by filtering any k-itemset out of the hash table, if the hash entry does not have minimum support. Partitioning technique [16] divides the transactional database into *n* non-overlapping partitions. It requires two database scans. For each partition, frequent itemsets are found during the first scan and they are called as local frequent itemsets. A local frequent itemset may or may not be frequent with respect to the entire database. The collection of all local frequent itemsets forms the global candidate itemsets. In the second scan, these global candidate itemsets are used to find the global frequent items using the entire database. A sampling approach [18] picks a random sample of transactions *S* from the database, and then search for the frequent itemsets in *S* instead of the total database. A Dynamic Itemset Counting (DIC) [7] method partitions a database into blocks marked by start points. New candidate itemsets can be added at any start point, unlike in Apriori, which determines new candidate itemsets immediately before each complete database scan. This technique is dynamic since it calculates support of all of the itemsets that have been counted so far, adding new candidate itemsets if all of their subsets are estimated to be frequent.

FP-growth [11], generate frequent itemsets without candidate generation. It works in a divide-and-conquer way. The first scan of the database derives a list of frequent items in which items are ordered by frequency-descending order. According to the frequency-descending list, the database is compressed into a Frequent-Pattern tree (FP-tree), which retains the itemset association information. The FP-tree is mined by starting from each frequent length-1 pattern (as an initial suffix pattern), constructing its conditional pattern base, then constructing its conditional FP-tree, and performing mining recursively such a tree. The pattern growth is achieved by the concatenation of the suffix pattern with the frequent patterns generated from a conditional FP-tree.

Both the Apriori and FP-growth methods mine frequent patterns from a set of transactions in horizontal data format (i.e., {TID: itemset}), where TID is a transaction-id and itemset is the set of items bought in transaction TID. Alternatively, mining can also be performed with data presented in vertical data format (i.e., {item: TID_set}).

Zaki proposed Equivalence CLASS Transformation (Eclat) [23] algorithm by exploring the vertical data format. The first scan of the database builds the TID_set of each single item. Starting with a single item (k=1), the frequent (k+1)-itemsets grown from a previous k-itemset can be generated according to the Apriori property, with a depth-first computation order similar to FP-growth. The computation is done by intersection of the TID_sets of the frequent k-itemsets to compute the TID_sets of the corresponding (k+1)-itemsets. This process repeats, until no frequent itemsets or no candidate itemsets can be found. Besides taking advantage of the Apriori property in the generation of candidate (k+1)-itemset from frequent k-itemsets, another merit of this method is that there is no need to scan the database to find the support of (k+1)-itemsets (for k≥1). This is because the TID_set of each k-itemset carries the complete information required for counting such support.

A major challenge in mining frequent patterns from a large data set is the fact that such mining often generates a huge number of patterns satisfying the min_sup threshold, especially when min_sup is set low. This is because if a pattern is frequent, each of its sub patterns is frequent as well. A large pattern will contain an exponential number of smaller, frequent sub-patterns. To overcome this problem closed and maximal frequent patterns were proposed. Pasquier et al. proposed A-Close [15] for mining closed frequent patterns and other closed pattern mining algorithms include CLOSET [14], CHARM [22], CLOSET+ [21], and FPClose [9]. The main challenge in closed (maximal) frequent pattern mining is to check whether a pattern is closed (maximal). There are two strategies to approach this issue: (1) to keep track of the TID list of a pattern and index the pattern by hashing its TID values. This method is used by CHARM which maintains a compact TID list called a diffset and (2) to maintain the discovered patterns in a pattern-tree similar to FP-tree. This method is exploited by CLOSET+ and FPClose.

Mining maximal patterns was first studied by Bayardo [5], where MaxMiner, an Apriori-based, level-wise, breadth-first search method was proposed to find max-itemset by performing superset frequency pruning and subset infrequency pruning for search space reduction. Another efficient method called MAFIA, proposed by Burdick et al. [8], uses vertical bitmaps to compress the transaction id list, thus improving the counting efficiency.

The above algorithms are not considering the time point at which the transaction occurs. So the dynamic behavior of the frequent pattern is not revealed. To overcome this problem Wan et. al.[19] [20], proposed new patterns called transitional patterns and TP-Mine algorithm to mine transitional patterns.

## 3. PRELIMINARIES AND DEFINITIONS

Mining frequent patterns is one of the fundamental operations in data mining applications for extracting interesting patterns from databases. Let I= $\{i_1, i_2 \dots i_n\}$ be a set of items. Let D be a set of database transactions where each transaction T is a set of items and ||D|| be the number of transactions in D. Given X= $\{i_j \dots i_k\} \subseteq$ I (j ≤k and 1 ≤ j, k ≤ n) is called a pattern. The support of a pattern X in D is the number of transactions in D that contains X. Pattern X will be called frequent if its support is no less than a user specified minimum support threshold.

*Definition 2.1:* The cover of an itemset X in D, denoted by *cov(X,D)*, is the number of transactions in which the item X appears.

*Definition 2.2:* An itemset X in a transaction database D has a support, denoted by *sup(X, D)*, which is the ratio between *cov(X, D)* to the number of transactions in D i.e., ||D||.

$$\sup(X, D) = \frac{cov(X,D)}{||D||}$$

*Definition 2.3:* Assuming that the transactions in a transaction database D are ordered by their time-stamps, the position of a transaction T in D, denoted by ρ(T), is the number of transactions whose time-stamp is less than or equal to that of T. Thus 1 ≤ ρ(T) ≤ ||D||.

*Definition 2.4:* The $i^{th}$ transaction of a pattern X in D, denoted by $T^i(X)$, is the $i^{th}$ transaction in *cov(X)* with transactions ordered by their positions, where 1≤ *i* ≤ *cov(X, D)*.

*Definition 2.5:* The $i^{th}$ milestone of a pattern X in D, denoted by $\xi^i(X)$, is defined as

$$\xi^i(X) = \frac{\rho(Ti(X))}{||D||} \times 100\% \quad \text{where } 1 \leq i \leq cov(X).$$

*Definition 2.6:* The support of the pattern X before its $i^{th}$ milestone in D, denoted by $sup_-^i(X)$ is defined as

$$sup_-^i(X) = \frac{i}{\rho(Ti(X))} \quad \text{where } 1 \leq i \leq cov(X).$$

*Definition 2.7:* The support of the pattern X after its $i^{th}$ milestone in D, denoted by $sup_+^i(X)$ is defined as

$$sup_+^i(X) = \frac{cov(x) - i}{||D|| - \rho(Ti(X))} \quad \text{where } 1 \leq i \leq cov(X).$$

*Definition 2.8:* The Transitional ratio of the pattern X at its $i^{th}$ milestone in D is defined as

$$tran^i(X) = \frac{sup_+^i(X) - sup_-^i(X)}{Max(sup_+^i(X), sup_-^i(X))} \quad \text{where } 1 \leq i \leq cov(X).$$

*Definition 2.9:* A pattern X is a transitional pattern (TP) in D if there exist at least one milestone of X, $\xi^k(X) \in T_\xi$, such that:

$$1.\ sup_-^k(X) \geq t_s \text{ and } sup_+^k(X) \geq t_s \text{ and } \quad 2.\ |tran^i(X)| \geq t_t,$$

*where $T_\xi$ is a range of $\xi^i(X)$ ($1 \leq i \leq cov(X)$), $t_s$ and $t_t$ are called pattern support threshold and transitional pattern threshold, respectively. X is called a Positive Transitional Pattern (PTP)* *when* $tran^k(X)>0$; *and X is called a Negative Transitional Pattern (NTP) when* $tran^k(X)<0$.

*Definition 2.10:* The significant frequency-ascending milestone (SFAM) of a positive transitional pattern X with respect to a time period $T_\xi$ is defined as a tuple, $(\xi^M(X), tran^M(X))$, where $\xi^M(X) \in T_\xi$ is the $M^{th}$ milestone of X such that

$$1.\ sup_{-}^{M}(X) \geq t_s;\ \text{ and }\ 2.\ \forall \xi^i(X) \in T_\xi, tran^M(X) \geq tran^i(X)$$

*Definition 2.11:* The significant frequency-descending milestone (SFDM) of a negative transitional pattern X with respect to a time period $T_\xi$ is defined as a tuple, $(\xi^N(X), tran^N(X))$, where $\xi^N(X) \in T_\xi$ is the $N^{th}$ milestone of X such that

$$1.\ sup_{+}^{N}(X) \geq t_s;\ and\ 2.\ \forall \xi^i(X) \in T_\xi, tran^N(X) \leq tran^i(X)$$

# 4. TP-MINE ALGORITHM

Consider the following transaction database shown in Table 1.

Table 1: Example Database

| TID | List of Item IDs | Time stamp |
|---|---|---|
| 001 | P1,P2,P3,P5 | Nov, 2005 |
| 002 | P1,P2 | Dec, 2005 |
| 003 | P1,P2,P3,P8 | Jan, 2006 |
| 004 | P1,P2,P5 | Feb, 2006 |
| 005 | P1,P2,P4 | Mar, 2006 |
| 006 | P1,P2,P4,P5,P6 | Apr, 2006 |
| 007 | P1,P2,P3,P4,P6 | May, 2006 |
| 008 | P1,P4,P6 | Jun, 2006 |
| 009 | P4,P5,P6 | Jul, 2006 |
| 010 | P1,P2,P3,P4,P5,P6 | Aug, 2006 |
| 011 | P1,P3,P4,P6 | Sep, 2006 |
| 012 | P1,P3,P5 | Oct, 2006 |
| 013 | P1,P2,P3,P6,P7 | Nov, 2006 |
| 014 | P1,P3,P4,P5 | Dec, 2006 |
| 015 | P1,P3,P4 | Jan, 2007 |
| 016 | P1,P2,P3,P5 | Feb, 2007 |

TP-Mine algorithm [20] generates the set of positive and negative transitional patterns with their significant milestones.
**Input**: A transaction database (D), milestone range ($T_\xi$), pattern support threshold ($t_s$), and transitional pattern ($t_t$).

We considered D as set of transactions listed in Table 1, $T_\xi$= {25%, 75%}, $t_s$=0.05, $t_t$=0.5

This algorithm has two major phases. In the first phase all frequent itemsets along with their support counts are generated using Apriori [3] or FP-growth [11] with $t_s$ as minimum support threshold. In this step, the algorithm generates *n* number of frequent patterns whose support ≥ $t_s$ from the transaction database D. The resultant frequent patterns (n=87) are shown in Table 2.

Table 2: Set of frequent patterns

| FP | Sup | FP | Sup | FP | Sup | FP | Sup |
|---|---|---|---|---|---|---|---|
| P1 | 15 | P3,P5 | 5 | P1,P6,P7 | 1 | P1,P2,P5,P6 | 2 |
| P2 | 10 | P3,P6 | 4 | P2,P3,P4 | 2 | P1,P2,P6,P7 | 1 |
| P3 | 10 | P3,P7 | 1 | P2,P3,P5 | 3 | P1,P3,P4,P5 | 2 |
| P4 | 9 | P3,P8 | 1 | P2,P3,P6 | 3 | P1,P3,P4,P6 | 3 |
| P5 | 8 | P4,P5 | 4 | P2,P3,P7 | 1 | P1,P3,P5,P6 | 1 |
| P6 | 7 | P4,P6 | 6 | P2,P3,P8 | 1 | P1,P3,P6,P7 | 1 |
| P7 | 1 | P5,P6 | 3 | P2,P4,P5 | 2 | P1,P4,P5,P6 | 2 |
| P8 | 1 | P6,P7 | 1 | P2,P4,P6 | 3 | P2,P3,P4,P5 | 1 |
| P1,P2 | 10 | P1,P2,P3 | 6 | P2,P5,P6 | 2 | P2,P3,P4,P6 | 2 |
| P1,P3 | 10 | P1,P2,P4 | 4 | P2,P6,P7 | 1 | P2,P3,P5,P6 | 1 |
| P1,P4 | 8 | P1,P2,P5 | 5 | P3,P4,P5 | 2 | P2,P3,P6,P7 | 1 |
| P1,P5 | 7 | P1,P2,P6 | 4 | P3,P4,P6 | 3 | P2,P4,P5,P6 | 2 |
| P1,P6 | 6 | P1,P2,P7 | 1 | P3,P5,P6 | 1 | P3,P4,P5,P6 | 1 |
| P1,P7 | 1 | P1,P2,P8 | 1 | P3,P6,P7 | 1 | P1,P2,P3,P4,P5 | 1 |
| P1,P8 | 1 | P1,P3,P4 | 5 | P4,P5,P6 | 3 | P1,P2,P3,P4,P6 | 2 |
| P2,P3 | 6 | P1,P3,P5 | 5 | P1,P2,P3,P4 | 2 | P1,P2,P3,P5,P6 | 1 |
| P2,P4 | 4 | P1,P3,P6 | 4 | P1,P2,P3,P5 | 3 | P1,P2,P3,P6,P7 | 1 |
| P2,P5 | 5 | P1,P3,P7 | 1 | P1,P2,P3,P6 | 3 | P1,P2,P4,P5,P6 | 2 |
| P2,P6 | 4 | P1,P3,P8 | 1 | P1,P2,P3,P7 | 1 | P1,P3,P4,P5,P6 | 1 |
| P2,P7 | 1 | P1,P4,P5 | 3 | P1,P2,P3,P8 | 1 | P2,P3,P4,P5,P6 | 1 |
| P2,P8 | 1 | P1,P4,P6 | 5 | P1,P2,P4,P5 | 2 | P1,P2,P3,P4,P5,P6 | 1 |
| P3,P4 | 5 | P1,P5,P6 | 2 | P1,P2,P4,P6 | 3 | | |

In the second phase, the algorithm computes the support counts ($c_k$) of all the frequent patterns on the set from the first transaction to the transaction just before the time period $T_\xi$. Then it computes the milestones of $P_k$ ($1 \leq k \leq n$) within the range $T_\xi$. At each valid milestone $\xi^{c_k}(P_k)$ during the scan, it calculates the support of $P_k$ before $\xi^{c_k}(P_k)$ i.e., $sup_-^{c_k}(P_k)$ and support of $P_k$ after $\xi^{c_k}(P_k)$ i.e., $sup_+^{c_k}(P_k)$. If both of them are greater than $t_s$, the algorithm then checks the transitional ratio of $P_k$. If the ratio is greater than $t_t$ then $P_k$ is a positive transitional pattern. Then the algorithm checks whether the transitional ratio of $P_k$ is greater than the current maximal transitional ratio of $P_k$. If yes, the set of significant frequency ascending milestones of $P_k$ is set to contain $\{\xi^{c_k}(P_k), tran^{c_k}(P_k)\}$ as its single element. If not but it is equal to the current maximal transitional ratio of $P_k$, $\{\xi^{c_k}(P_k), tran^{c_k}(P_k)\}$ is added into significant frequency ascending milestone of $P_k$. Similarly it computes negative transitional patterns and their significant milestones if $tran^{c_k}(P_k) \leq -t_t$. The SFAM and SFDM are shown in table 3 and table 4 respectively.

Table 3: Set of Positive transitional patterns and their ascending milestones.

| Pattern | Ascending Milestone | Pattern | Ascending Milestone |
|---|---|---|---|
| P3 | aug2006(62.5%), 60.000 % | P3,P5 | aug2006(62.5%), 60.000 % |
| P4 | mar2006(31.25%), 72.500 % | P3,P6 | may2006(43.75%), 57.143 % |
| P6 | apr2006(37.5%), 72.222 % | P4,P6 | apr2006(37.5%), 66.667 % |
| P1,P3 | aug2006(62.5%), 60.000 % | P1,P3,P4 | may2006(43.75%), 67.857 % |
| P1,P4 | mar2006(31.25%), 68.571 % | P1,P3,P5 | aug2006(62.5%), 60.000 % |
| P1,P6 | apr2006(37.5%), 66.667 % | P1,P3,P6 | may2006(43.75%), 57.143 % |
| P3,P4 | may2006(43.75%), 67.857 % | P1,P4,P6 | apr2006(37.5%), 58.333 % |

Table 4: Set of Negative transitional patterns and their descending milestones.

| Pattern | Descending Milestone | Pattern | Descending Milestone |
|---|---|---|---|
| P2 | may2006(43.75%), -66.667 % | P4,P6 | aug2006(62.5%), -66.667 % |
| P6 | sep2006(68.75%), -63.333 % | P1,P2,P4 | may2006(43.75%), -74.074 % |
| P1,P2 | may2006(43.75%), -66.667 % | P1,P2,P5 | apr2006(37.5%), -60.000 % |
| P1,P6 | sep2006(68.75%), -56.000 % | P1,P4,P6 | aug2006(62.5%), -58.333 % |
| P2,P4 | may2006(43.75%), -74.074 % | P2,P4,P6 | may2006(43.75%), -61.111 % |
| P2,P5 | apr2006(37.5%), -60.000 % | P1,P2,P4,P6 | may2006(43.75%), -61.111 % |

It was observed that the frequent patterns which contains items P7 or P8 (i.e., 24 patterns) are not used in the second phase of the algorithm, because these items are not contained in the transactions whose position satisfying $T_{\xi}$. Therefore generating these 24 patterns in the first phase of the algorithm is useless. So, we are proposing a method for pruning these useless frequent patterns in the first phase of the algorithm.

## 5. ETP-MINE (EFFICIENT TP-MINE) ALGORITHM

In this section, we are presenting an Efficient TP-Mine algorithm which eliminates the useless frequent patterns. The algorithm efficiently generates positive and negative transitional patterns and their significant milestones with respect to a pattern support threshold and transitional pattern threshold.

*Algorithm: ETP-Mine*

*Input:*
A transaction database (D), an appropriate milestone range that the user is interested ($T_{\xi}$), pattern support threshold ($t_s$), and transitional pattern threshold ($t_t$).

*Output:*
The set of transitional patterns ($S_{PTP}$ and $S_{NTP}$) with their significant milestones.

*Method:*

1: Generate frequent 1-itemsets ($L_1$) and their supports from D using the candidate 1-itemsets which appears in the transactions whose position satisfying $T_\xi$
2: Extract frequent patterns, $P_1, P_{2,} P_3, \ldots, P_n$, and their supports using $L_1$ with min_sup $= t_s$.
3: Scan the transactions from the first transaction to the last transaction before $T_\xi$ to compute the support counts, $c_k$ (1 ≥k ≥ n), of all the n frequent patterns on this part of the database.
4: $S_{PTP}$ =ø, $S_{NTP}$=ø
5: for all k = 1 to n do
6: MaxTran ($P_k$) = 0, MinTran ($P_k$) = 0
7: $S_{FAM}$ ($P_k$) =ø, $S_{FDM}$ ($P_k$) =ø
8: end for
9: for all transactions $T_i$ whose position satisfying $T_\xi$ do
10: for k = 1 to n do
11: if $T_i \supseteq P_k$ then

$c_k = c_k + 1$
if $sup_{-}^{c_k}(P_k) \geq t_s$ and $sup_{+}^{c_k}(P_k) \geq t_s$ then
if $tran^{c_k}(P_k) \geq t_t$ then
if $P_k \notin S_{PTP}$ then
Add $P_k$ to $S_{PTP}$
end if
if $tran^{c_k}(P_k) >$ MaxTran $(P_k)$ then
$S_{FAM}(P_k) = \{\xi^{c_k}(P_k), tran^{c_k}(P_k)\}$
MaxTran $(P_k) = tran^{c_k}(P_k)$
else if $tran^{c_k}(P_k) =$ MaxTran $(P_k)$ then
Add $\{\xi^{c_k}(P_k), tran^{c_k}(P_k)\}$ to $S_{FAM}(P_k)$
end if
else if $tran^{c_k}(P_k) \leq -t_t$ then
if $P_k \notin S_{NTP}$ then
Add $P_k$ to $S_{NTP}$
end if
if $tran^{c_k}(P_k) <$ MinTran $(P_k)$ then
$S_{FDM}(P_k) = \{\xi^{c_k}(P_k), tran^{c_k}(P_k)\}$
MinTran $(P_k) = tran^{c_k}(P_k)$
else if $tran^{c_k}(P_k) =$ MinTran $(P_k)$ then
Add $\{\xi^{c_k}(P_k), tran^{c_k}(P_k)\}$ to $S_{FDM}(P_k)$
end if
end if
end if
end if
end for
end for
return $S_{PTP}$ and $S_{FAM}(P_k)$ for each $P_k \in S_{PTP}$
return $S_{NTP}$ and $S_{FDM}(P_k)$ for each $P_k \in S_{NTP}$

By executing this algorithm on the transaction database given in the Table 1, the number of frequent patterns generated is reduced to 63, which are shown in Table 5 and the positive and negative transitional patterns generated are same as shown in Table 3 and Table 4.

Table 5: Frequent Patterns generated using ETP-Mine algorithm

| FP | Sup | FP | Sup | FP | Sup | FP | Sup | FP | Sup |
|---|---|---|---|---|---|---|---|---|---|
| P1 | 15 | P2,P5 | 5 | P1,P3,P5 | 5 | P3,P5,P6 | 1 | P2,P3,P4,P6 | 2 |
| P2 | 10 | P2,P6 | 4 | P1,P3,P6 | 4 | P4,P5,P6 | 3 | P2,P3,P5,P6 | 1 |
| P3 | 10 | P3,P4 | 5 | P1,P4,P5 | 3 | P1,P2,P3,P4 | 2 | P2,P4,P5,P6 | 2 |
| P4 | 9 | P3,P5 | 5 | P1,P4,P6 | 5 | P1,P2,P3,P5 | 3 | P3,P4,P5,P6 | 1 |
| P5 | 8 | P3,P6 | 4 | P1,P5,P6 | 2 | P1,P2,P3,P6 | 3 | P1,P2,P3,P4,P5 | 1 |
| P6 | 7 | P4,P5 | 4 | P2,P3,P4 | 2 | P1,P2,P4,P5 | 2 | P1,P2,P3,P4,P6 | 2 |
| P1,P2 | 10 | P4,P6 | 6 | P2,P3,P5 | 3 | P1,P2,P4,P6 | 3 | P1,P2,P3,P5,P6 | 1 |
| P1,P3 | 10 | P5,P6 | 3 | P2,P3,P6 | 3 | P1,P2,P5,P6 | 2 | P1,P2,P4,P5,P6 | 2 |
| P1,P4 | 8 | P1,P2,P3 | 6 | P2,P4,P5 | 2 | P1,P3,P4,P5 | 2 | P1,P3,P4,P5,P6 | 1 |
| P1,P5 | 7 | P1,P2,P4 | 4 | P2,P4,P6 | 3 | P1,P3,P4,P6 | 3 | P2,P3,P4,P5,P6 | 1 |
| P1,P6 | 6 | P1,P2,P5 | 5 | P2,P5,P6 | 2 | P1,P3,P5,P6 | 1 | P1,P2,P3,P4,P5,P6 | 1 |
| P2,P3 | 6 | P1,P2,P6 | 4 | P3,P4,P5 | 2 | P1,P4,P5,P6 | 2 | | |
| P2,P4 | 4 | P1,P3,P4 | 5 | P3,P4,P6 | 3 | P2,P3,P4,P5 | 1 | | |

## 6. EXPERIMENTAL RESULTS

To demonstrate the efficiency of the ETP-Mine algorithm we have done experiments on synthetic data obtained from a Grocery shop with 25 transactions and 16 items. Table 6 and figure 1. shows the comparison of TP-Mine algorithm and ETP-Mine algorithm against the Grocery data. Our experimental results showed that the proposed algorithm is highly efficient and scalable.

Table 6: Comparison of TP-Mine and ETP-Mine algorithms.

| | Number of frequent patterns generated in | |
|---|---|---|
| **Min_sup** | **TP-Mine** | **ETP-Mine** |
| 1 | 141 | 114 |
| 2 | 79 | 76 |
| 3 | 63 | 63 |

Figure1: Comparison of TP-Mine and ETP-Mine algorithms.

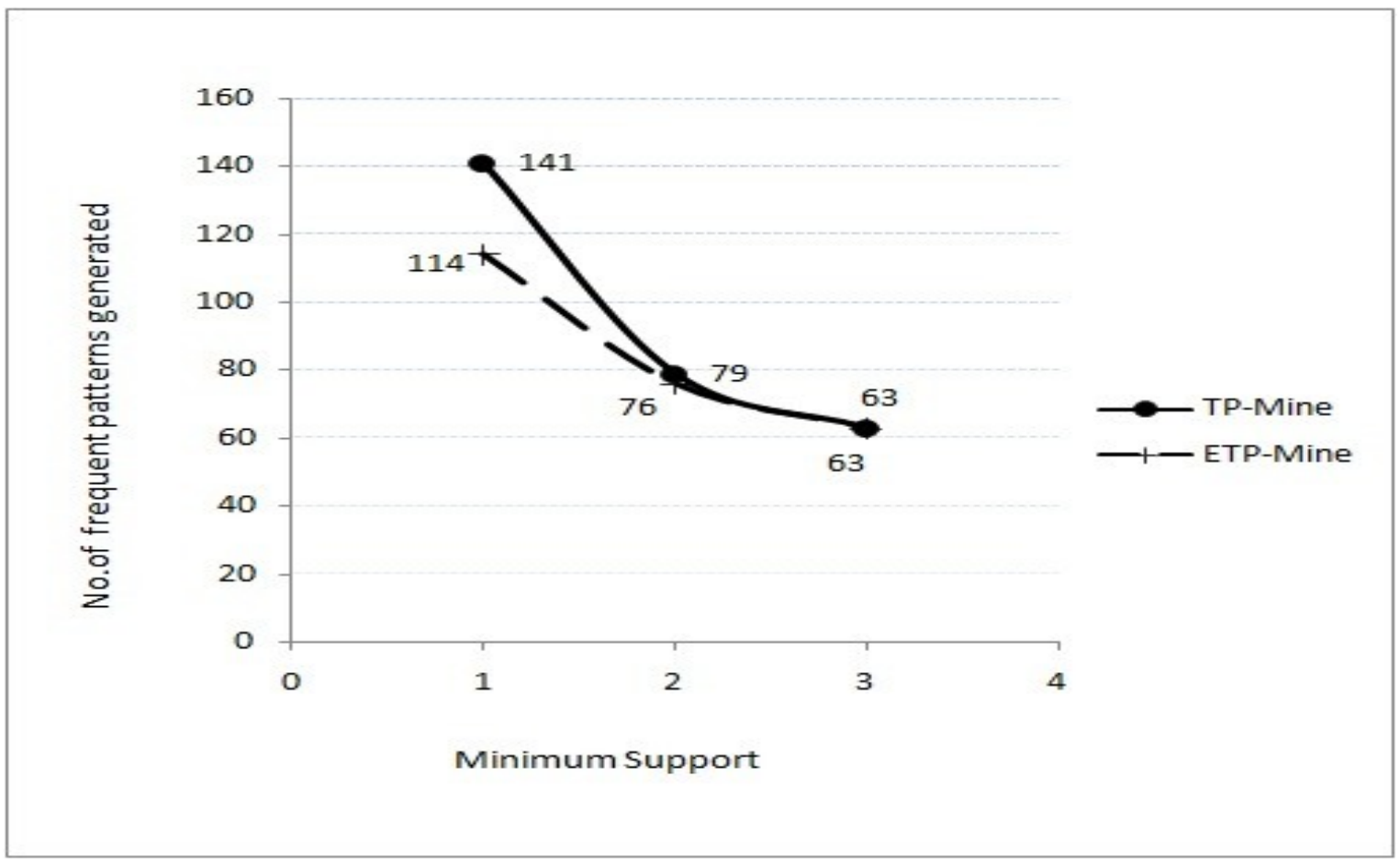

## 7. CONCLUSIONS

The existing TP-Mine algorithm generates more number of frequent patterns in the first phase, some of which are not useful in generating transitional patterns and their milestones. In this paper we presented a modification to the existing TP-Mine algorithm, which eliminates useless frequent patterns for generating transitional patterns. Our experimental results shows that, for the smaller values of minimum support threshold the number of frequent patterns generated are drastically reduced. So, our algorithm needs less number of comparisons in generating transitional patterns in the second phase. ETP-Mine algorithm achieves better efficiency and performance compared to TP-Mine algorithm when the database size and number of items increased.

B. Kiran Kumar received M.C.A. from Kakatiya University in 1998, and he is pursuing M.Tech in Computer Science and Engg. in JNTU, Hyderabad. He is working as Associate Professor in the department of M.C.A. He delivered guest lectures in the field of Data mining at various engineering colleges. His research interest includes Data mining. He is a member of ISTE.

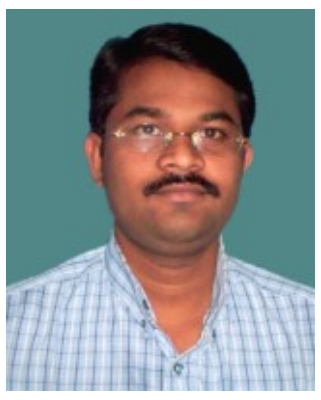

A. Bhaskar received M.C.A. from Kakatiya University in 1998, and he is pursuing M.Tech in Computer Science and Engg. in JNTU, Hyderabad. He is working as Associate Professor in the department of M.C.A. He delivered guest lectures in the field of Artificial Intelligence at various engineering colleges. His research interest includes Data mining. He is a member of ISTE, IAENG.

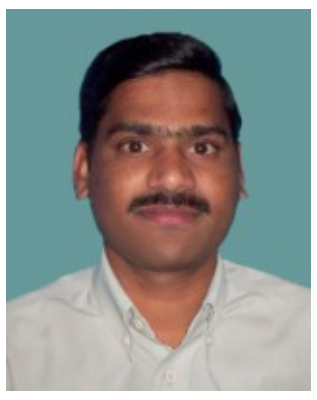